\documentclass[fleqn,twoside]{article}
\usepackage{espcrc2}

\usepackage{graphicx}
\usepackage[figuresright]{rotating}

\newcommand{\nn}{\nonumber}
\newcommand{\bd}{\begin{document}}
\newcommand{\ed}{\end{document}}
\newcommand{\bc}{\begin{center}}
\newcommand{\ec}{\end{center}}
\newcommand{\be}{\begin{eqnarray}}
\newcommand{\ee}{\end{eqnarray}}

\newcommand{\ttbs}{\char'134}
\newcommand{\AmS}{{\protect\the\textfont2
  A\kern-.1667em\lower.5ex\hbox{M}\kern-.125emS}}

\title{T violation in $\Lambda_{b}\rightarrow \Lambda  \ell^+
\ell^-$ decays }

\author{Chuan-Hung Chen\address{Institute of Physics, Academia Sinica\\
Taipei, Taiwan 115, Republic of China },
C.~Q.~Geng\address{Department of Physics, National Tsing Hua University\\
Hsinchu, Taiwan, Republic of China }
\thanks{Talk presented by C.Q. Geng
 at the 5th International Conference on Hyperons, Charm and Beauty 
Hadrons}
 and 
J.~N.~Ng
\address{Theory Group, TRIUMF, 4004 Wesbrrok Mall\\
Vancouver, B.C. V6T 2A3, Canada}
}
       
\begin{document}

\begin{abstract}
We examine the T-odd transverse lepton and $\Lambda$ polarizations
 in the baryonic decays of
$\Lambda_b\to\Lambda l^+l^-\ (l=e\,,\mu)$
to study the T violating effects. 
We show that the lepton polarizations
are suppressed but the $\Lambda$ ones
 can be as large as $50\%$ in CP violating
theories beyond the standard model such as the SUSY models,
which can be tested in various future hadron colliders.

\vspace{1pc}
\end{abstract}

\maketitle


Recently, time-reversal violation (TV) has been measured
experimentally in the $K^0$ system \cite{TVk}, and thus
complements the information on CP violation (CPV) that has been
steadily accumulating for the past thirty seven years. The data is
in accordance with the CPT theorem which is fundamental to local
quantum field theories with Lorentz invariance and the usual
spin-statistics connection. However, the origin of the violation
remains unclear. In the standard model, CPV or TV arises from a
unique physical phase in the Cabibbo-Kobayashi-Maskawa (CKM) quark
mixing matrix \cite{ckm}. This paradigm also predicts CPV effect
in the b-quark system. To test the accuracy of this paradigm and
to search for other sources of CPV one needs to look for new
processes.
Indeed this is an
important quest of the B-factories. In addition we deem it
particularly interesting if the time reversal symmetry violation
can be directly detected in the b-system rather than inferring it
as a consequence of CPT invariance. 
In this talk, we will present our recent 
studies on T violation
 in the baryonic decays of
$\Lambda_b\to\Lambda l^+l^-\ (l=e,\mu )$ \cite{geng1,geng2,geng3,geng4}.

 It is known that for a general three-body decay of a baryon the triple 
spin-momentum
correlations, such as $\vec{s}_k\cdot \left( \vec{p}_{i}\times \vec{p}%
_{j}\right) $, are T-odd observables, where $\vec{s}_k$ and
$\vec{p}_{i,j}$ are the spin and momentum vectors of the final
particles, respectively. There are a number of different sources
that might give rise to these T-odd observables. The most
important ones being the weak CPV such as the CKM phase of the SM.
 However, final state
interactions such as QCD for non-leptonic decays
 or the electromagnetic (EM) interaction
among the final state particles can also make contributions. These
are usually less interesting and they could even hide the signals
from the weak CPV. We note that the T-odd triple correlations do
not need non-zero strong phases unlike some of CP violating
observables, such as the rate asymmetry between a particle and its
antiparticle.
In $\Lambda _{b}\rightarrow \Lambda l^{+}l^{-}$ decays, 
we can
use either the polarization of the lepton ($s_{l}$) or the $\Lambda $ baryon ($s_{\Lambda}$)
to study the T-odd correlations. 
It is interesting to note that the decays can be used to analyze the 
helicity structure of the interaction, which is impossible in the B meson 
system.


We start with the effective Hamiltonian for $b\rightarrow sl^{+}l^{-}$ by
including the right-handed couplings in the hadronic sector are given by
\begin{eqnarray}
{\cal H}( b\rightarrow sl^{+}l^{-})
 =\frac{G_{F}\alpha _{em}}{%
\sqrt{2}\pi }V_{tb}V_{ts}^{*}(H_{1\mu }L_{V}^{\mu } +H_{2\mu
}L_{A}^{\mu })  \label{hameff}
\nn
\end{eqnarray}
\begin{eqnarray}
H_{1\mu } &=&\bar{s}\gamma ^{\mu }(C_{9}^{L}P_{L}+C_{9}^{R}P_{R}) b
\nn\\
&& -\frac{2m_{b}}{q^{2}}\bar{s}i\sigma
_{\mu \nu }q^{\nu }( C_{7}^{L}P_{R}+C_{7}^{R}P_{L}) b\,,
  \nonumber
\\
H_{2\mu } &=&\bar{s}\gamma ^{\mu }(
C_{10}^{L}P_{L}+C_{10}^{R}P_{R}) b \,, \nonumber \\
L_{V}^{\mu } &=&\bar{l}\gamma ^{\mu }l\,,\ \
L_{A}^{\mu } \;=\;\bar{l}\gamma ^{\mu }\gamma _{5}l\,,
  \label{hameff1}
\end{eqnarray}
where $C_{i}^{L}$ and $C_{i}^{R}\left( i=7,9,10\right) $ denote the
effective Wilson coefficients of left- and right-handed couplings,
respectively. In the standard model,
\begin{eqnarray}
C_{9}^{L}=C_{9}^{eff}\,, &&C_{9}^{R}=0\,,  \nonumber \\
C_{10}^{L}=C_{10}\,, &&C_{10}^{R}=0\,,  \nonumber \\
C_{7}^{L}=C_{7}^{eff}\,, &&C_{7}^{R}={\frac{m_{s}}{m_{b}}}C_{7}^{eff}\,,
\label{SMWC}
\end{eqnarray}
where $C_{9}^{eff}$, $C_{10}$, and $C_{7}^{eff}$ are the standard Wilson
coefficients \cite{Buras}.

To study the exclusive decays of $\Lambda_b\to\Lambda l^+l^-$, one needs to
know the form factors in the transition of $\Lambda _{b}(p_{\Lambda
_{b}})\rightarrow \Lambda ( p_{\Lambda })$, parametrized generally as
follows:
\begin{eqnarray}
&&\left\langle \Lambda \right| \bar{s}\ \Gamma _{\mu }\ b\left| \Lambda
_{b}\right\rangle 
=f_{1}^{(T)}\bar{u}_{\Lambda }\gamma _{\mu }u_{\Lambda
_{b}}
\nn\\
&+&f_{2}^{(T)}\bar{u}_{\Lambda }i\sigma _{\mu \nu }\ q^{\nu }u_{\Lambda
_{b}}+f_{3}^{(T)}q_{\mu }\bar{u}_{\Lambda }u_{\Lambda _{b}},  
\nonumber \\
&&\left\langle \Lambda \right| \bar{s}\ \Gamma _{\mu }\gamma _{5}\ b\left|
\Lambda _{b}\right\rangle
 =g_{1}^{(T)}\bar{u}_{\Lambda }\gamma _{\mu
}\gamma _{5}u_{\Lambda _{b}}
\nn\\
&+&g_{2}^{(T)}\bar{u}_{\Lambda }i\sigma _{\mu \nu
}\ q^{\nu }\gamma _{5}u_{\Lambda _{b}}+g_{3}^{(T)}q_{\mu }\bar{u}_{\Lambda
}\gamma _{5}u_{\Lambda _{b}}\,,  \label{atcq}
\end{eqnarray}
where $\Gamma_{\mu}=\gamma _{\mu }\ (i\sigma_{\mu\nu})$,
and $f_i^{(T)}$ and $g_i^{(T)}$ are the form factors of vector (tensor) and
axial-vector (axial-tensor) currents, respectively. In the heavy quark
effective theory (HQET) \cite{MR}, the form factors in Eq. (\ref{atcq}) can
be simplified by using
\begin{eqnarray}
&&
\left\langle \Lambda (p_{\Lambda })\right| \bar{s}\Gamma b\left| \Lambda
_{b}(p_{\Lambda _{b}})\right\rangle
\nn\\
& =&\bar{u}_{\Lambda }\left( F_{1}(q^{2})+%
\slash{\!\!\!{v}}F_{2}(q^{2})\right) \Gamma u_{\Lambda _{b}}\,,
\end{eqnarray}
where $\Gamma $ denotes the Dirac matrix, $v=p_{\Lambda _{b}}/M_{\Lambda
_{b}}$ is the four-velocity of $\Lambda _{b}$, and $q=p_{\Lambda
_{b}}-p_{\Lambda }$ is the momentum transfer, and the relations among the
form factors can be found in Ref. \cite{geng1}. Explicitly, under the
HEQT, we have
\begin{eqnarray}
f_{1} &=& g_{2} = f^{T}_{2}=g^{T}_{2} = F_{1}+\sqrt{r}F_{2}\,,
  \nonumber 
\\
\rho &\equiv &M_{\Lambda _{b}}\left( {\frac{f_{2}+g_{2}}{f_{1}+g_{1}}}%
\right) = \frac{M_{\Lambda _{b}}}{q^2}\left( {\frac{f^{T}_{1}+g^{T}_{1}}{f_{1}
+g_{1}}}\right)
\nn
\\
&& = {\frac{F_{2}}{F_{1}+\sqrt{r}F_{2}}}\,.  \label{HQETff}
\end{eqnarray}
 From experiments \cite{CLEO} and theories \cite{MR,Huang}, one finds that
$|F_1/F_2|>1$ and $|\rho|<1$.


To explore the T violating effects using  
the lepton or $\Lambda $ 
spin polarization, we write $i$ ($i=l$
or $\Lambda $) four-spin vector in 
terms 
of a unit
vector, $\hat{\xi}^i$, along the spin in its rest frame, as
\begin{eqnarray}
s_{i0}\,=\,\frac{\vec{p}_{i}\cdot \hat{\xi}_i}{M_{i}},\qquad 
\vec{s}_i\,=\,\hat{\xi}_i+\frac{s_{i0}}{E_{i}+M_{i }}\vec{p}_{i},
\end{eqnarray}
and choose the unit vectors along the longitudinal, normal, transverse
components of the $i$ polarization, to be
\begin{eqnarray}
\hat{e}^i_{L} &=&\frac{\vec{p}_{i}}{|\vec{p}_{i}| },\
\hat{e}^i_{N} \;=\;\frac{\vec{p}_{i}\times \left(\vec{p}_{l^{-}}\times
\vec{p}_{\Lambda}\right) }{\left| \vec{p}_{i }\times \left( \vec{p}%
_{l^{-}}\times \vec{p}_{\Lambda }\right) \right| },  \nonumber \\
\hat{e}^i_{T} &=&\frac{\vec{p}_{l^{-}}\times \vec{p}_{\Lambda }}{\left| 
\vec{p}%
_{l^{-}}\times \vec{p}_{\Lambda }\right| }\,,  \label{uv}
\end{eqnarray}
respectively. Hence, the differential decay rates with polarized
$i$ is given by
\begin{eqnarray}
d\Gamma^i(t) &=&\frac{1}{2}d\Gamma ^{0}(t)\left[ 1+\vec{P}^i\cdot 
\hat{\xi}^i\right]\,,
\label{diffrate} 
\end{eqnarray}
where $t=E_{\Lambda }/M_{\Lambda _{b}}$,
$\vec{P}^i$ is the $i=l$ or $\Lambda$ polarization vector, defined 
by
\begin{eqnarray}
\vec{P}^i=P^i_{L}\hat{e}^i_{L}+P^i_{N}\hat{e}^i_{N}+P^i_{T}\hat{e}^i_{T}\,,
\end{eqnarray}
and 
$d\Gamma ^{0}(t)$
is given in Ref. \cite{geng4}.
 With the T odd transverse $i $ polarization defined by
 \be\nn 
P_{T}^i&=&\frac{d\Gamma^i \left( \hat{\xi}_i\cdot
\hat{e}_T^{i}=1\right) -d\Gamma^i \left( \hat{\xi}_i\cdot
\hat{e}_T^{i}=-1\right) }{d\Gamma^i \left( \hat{\xi}_i\cdot 
\hat{e}_T^{i}=1\right) +d\Gamma^i \left( \hat{\xi}_i\cdot
\hat{e}_T^{i}=-1\right) },
 \nn\ee
 we obtain
\be\nn 
P_T^l &\propto& {m_l\over M_{\Lambda_b}}
Im\left(C^L_9C^{L*}_{10}+C^R_9C^{R*}_{10}\right)\,,\ \
\\ 
&&{m_l\over
M_{\Lambda_b}}Im\left(C^L_7C^{L*}_{10}+C^R_7C^{R*}_{10}\right)
\label{Ptl}
\\   
\nn \\ \nn P_T^{\Lambda}&\propto&
Im\left(C^R_9C^{L*}_{10}-C^L_9C^{R*}_{10}\right)\,,\
\\ 
&&Im\left(C^L_7C^{R*}_{10}-C^R_7C^{L*}_{10}\right)
\label{PtL}
 \ee
It is easy to see that from Eq. (\ref{Ptl}) $P_T^l$ is suppressed since
it is always associated with the small lepton mass and
to have a non-zero value of $P_T^\Lambda$, 
from   Eq. (\ref{PtL}), 
it is necessary to have conditions of (a) the existence of $C_k^R$
and (b) a phase of $C_i^LC_j^R\ (i\neq j)$.
We remark that these conditions for $P_T^\Lambda$
are clearly different from those of the T odd transverse lepton
polarizations in both inclusive decay of $b\to s l^+l^-$
\cite{Tin} and exclusive ones, such as $B\to K^{(*)}l^+l^-$
\cite{TexM} and $\Lambda_b\to\Lambda l^+l^-$ \cite{geng2,geng3}.
Such property
distinguishes the T odd transverse $\Lambda$ polarizations from various
other T odd observables. 

In the standard model, since there are no $C^R_{9}$
and $C^R_{10}$ as seen from Eq. (\ref{SMWC}), we have
\be
P_T^{l}&\propto&
{\frac{m_l}{m_b}}Im\left(C_7^{eff}C_{10}\right)
\nn\\
P_T^{\Lambda}&\propto&
{\frac{m_s}{m_b}}Im\left(C_7^{eff}C_{10}\right)
\ee
which are all suppressed. 
We note that the contribution to $P_T^i$
from the EM final state interaction are $<O(10^{-3})$. 
Moreover, the long-distance (LD)
effects in the one-loop matrix elements of
$O_{1,2}$ \cite{Buras} and $\Lambda_b \rightarrow \Lambda J/\Psi$
with $J/\Psi\rightarrow l^+ l^-$
are absorbed to $C^{L}_{9}$. On the
other hand, it is clear that a large value of $P_T^\Lambda$, according to
Eq. (\ref{PtL}), can be obtained if a theory contains $C_9^R$ or
$C_{10}^R$ or a large $C_7^R$, with a non-zero phase.
Many theories beyond the
standard model could give rise to $C_9^R$ or $C_{10}^R$; examples are
the left-right symmetric and supersymmetric models.

\begin{figure}[htb]
\includegraphics[scale=0.5]{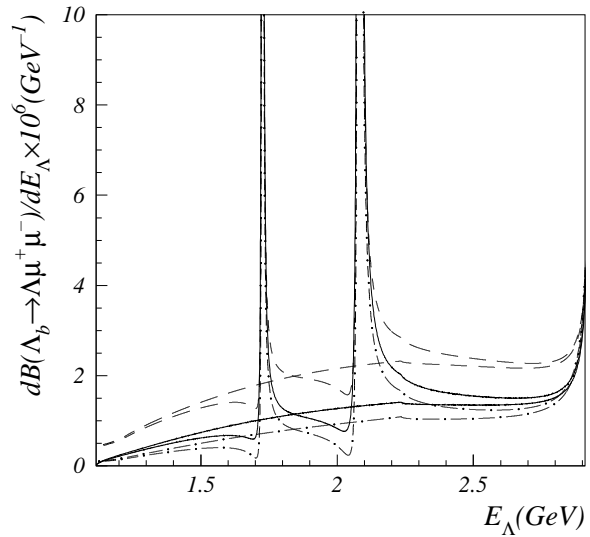}
\vspace{9pt}
\caption{
Differential BRs of $\Lambda_b\to\Lambda \mu^+\mu^-$ as
a function of $E_{\Lambda}$ with and without resonant shapes,
where the solid, dash-dotted and dashed curves stand for the
results of the standard and SUSY with and without R-parity models,
respectively. } \label{rpmurate}
\end{figure}

\begin{figure}[htb]
\includegraphics[scale=0.5]{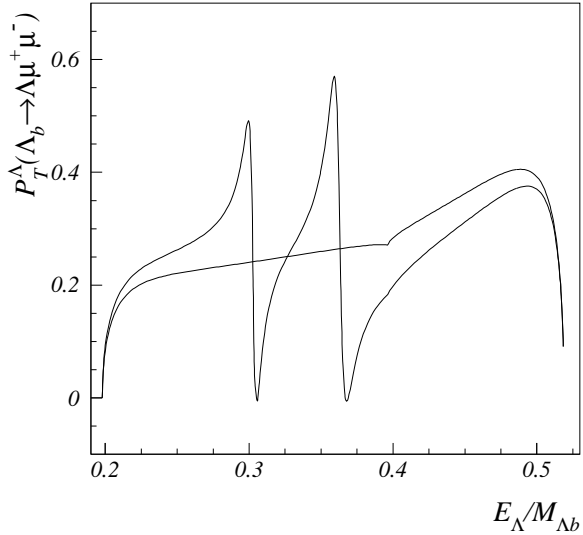}
\vspace{9pt}
\caption{
 Transverse $\Lambda$ polarization in $\Lambda_b\to\Lambda
\mu^+\mu^-$ as a function of $E_{\Lambda}/M_{\Lambda_b}$ in the SUSY model
with R-parity. The curves with and without resonant shapes represent
including and no long-distance contributions, respectively. }
\label{susypt}
\end{figure}

\begin{figure}[htb]
\includegraphics[scale=0.5]{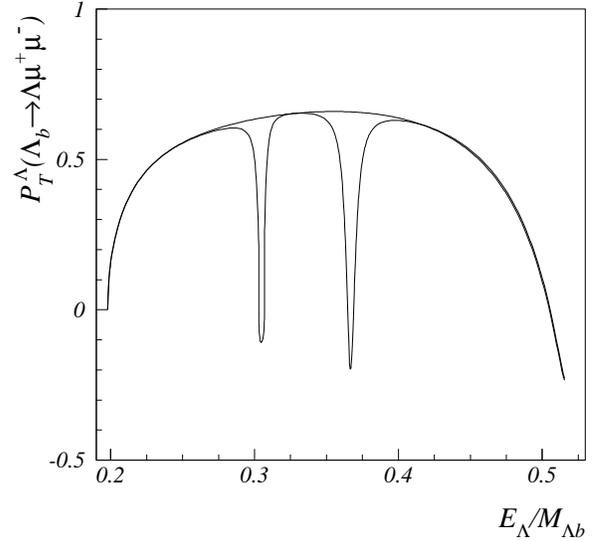}
\vspace{9pt}
\caption{
Same as Figure \ref{susypt}
but for the SUSY model without R-parity. } \label{rppt}
\end{figure}

To illustrate our result, we use
SUSY models 
\cite{Masiero,Nardi}
both with
and without R-parity as shown in Ref. \cite{geng4}.
In Figure \ref{rpmurate},
we give the differential branching ratios (BRs) of
$\Lambda_b\to\Lambda\mu^+ \mu^-$ with respect to $E_{\Lambda}$
 with and without including
resonant states of $\Psi$ and $\Psi ^{\prime }$,
and we find that
the integrated BRs for the latter are $(2.10,\,1.66,\,4.41)\times 10^{-6}$
for the standard and SUSY with
and without R-parity models, respectively.
In Figures \ref{susypt} and \ref{rppt}, 
we show
$P_T^\Lambda (\Lambda_b\to\Lambda \mu^+\mu^-)$ as a function of $%
E_{\Lambda}/M_{\Lambda_b}$
for the two types of models, with and without the LD contributions,
respectively. As seen
from the figures, even though the derivations of BRs to the standard
model result are insignificant, the transverse $\Lambda$ polarization
asymmetries can be over $50\%$ in both SUSY models with and without
R-parity. Similar results are also expected for the decay of
$\Lambda_b\to\Lambda e^+e^-$.
We remark that measuring a large $P_T^\Lambda$ in $\Lambda_b\to\Lambda
l^+l^-$ is a clean indication of T violation as well as new CP
violation mechanism beyond the standard model.
To measure $P_T^\Lambda (\Lambda_b\to\Lambda
\mu^+\mu^-)\sim 10\%$ at $3\sigma$ level, at least $4.5\times 10^7$ $%
\Lambda_b$ decays are required if we use $BR(\Lambda_b\to\Lambda
\mu^+\mu^-)\sim 2\times 10^{-6}$.
Clearly, the measurement could be done in the second generation
of B-physics experiments, such as LHCb, ATLAS, and CMS at the LHC, and
BTeV at
the Tevatron, which produce $\sim 10^{12}b\bar{b}$ pairs per year \cite{BB}.
This is certainly within reach of a super B factory under discussion now
\cite{SuperB}.
 Finally, we note
that similar results are expected for using polarized $\Lambda_b$ instead 
of $\Lambda$.

\noindent {\bf Acknowledgments}
This work was supported in part by
 the National Center for Theoretical Science,
National Science Council of the Republic of China under
 Contract Nos. NSC-90-2112-M-001-069 and NSC-90-2112-M-007-040,
 and National Science and Engineering Research Council of Canada.

\end{document}